\documentclass[11pt]{article}

\usepackage{graphicx}  
\title{The Possible Detection of Dark Energy on Earth Using Atom Interferometry}

\author{Martin L. Perl \\  SLAC National Accelerator Laboratory\\2575 Sand Hill Road, Menlo Park, California, 94025\\martin@slac.stanford.edu}

\begin{document}

\maketitle

\begin{abstract}
This paper describes the concept and the beginning of an experimental investigation of whether
it is possible to directly detect dark energy density on earth using atom interferometry. The concept is to null out the gravitational force using a double interferometer. This research provides a non-astronomical path for research on dark energy. The application of this method to other hypothetical weak forces and fields is also discussed. In the the final section I discuss the advantages of carrying out a dark energy density search in a satellite in earth orbit where more precise nulling of gravitational forces can be achieved. 
\end{abstract}

\section{Table Of Contents}

\begin{enumerate}
\item History of concept of using atom interferometry to investigate dark energy
\item Conventional beliefs about the nature of dark energy
\item Comparison of dark energy density with energy density of a weak electric field
\item The terrestrial gravitational force field and a possible dark energy force
\item Preliminary considerations on how well we can null out g
\item Our assumptions about the properties of dark energy that make the experiment
feasible 
\item Brief description of experimental search method
\item Nature of sought signal
\item Other very weak forces and fields
\item This experimental search method using an earth orbit satellite.
\end{enumerate}

\section{Origin of paper}
This paper is a summary of the talk I presented at the Les Rencontres de Physique de la Vallée d'Aoste, Results and Perspectives in Particle Physics, March 6, 2010

\section{History of concept of using atom interferometry to investigate dark energy}
The majority of astronomers and physicists accept the reality of dark energy but also
believe it can only be studied indirectly through observation of the structure and 
motions of galaxies. Astronomical investigation of dark energy are limited by their nature to:
 
\begin{itemize}
\item{Measurement of the dark energy density, $\rho_{DE}$.}
\item{Search for gross variations of $\rho_{DE}$ in the visible universe.}
\item{Elucidation of the change in $\rho_{DE}$ in the past.}
\item{There is no known way to investigate the nature of dark energy using observational astronomy.}
\end{itemize}

Several years ago \cite{perl1} I began to search for non-astronomical ways to investigate the nature of dark energy and realized that there was a possibility, albeit experimently speculative, to use atom interferometry \cite{perl2}. Atom interferometry is a research technology whose practice is about three decades old \cite{atom}.

I was then joined in this research area by Holger Mueller of the Physics Department, University of California at Berkeley and we continue to work together \cite{perl2}. This present paper recapitulates that paper\cite{perl2} in Secs. 1-9 and discusses three important new aspects of this research in Secs. 10-12.

\begin{itemize}
\item {The character of our signal is noiselike because of the motion of the earth through space, Sec. 10.}
\item{The research method is applicable to searches for hypothetical very weak forces and fields, Sec.11.}
\item{There are substantial advantages to eventually carying out these searches in a satellite in an earth orbit, Sec.12.}
\end{itemize}

We use MKS units rather than astronomical units to emphasize practical laboratory experimental designs and considerations. Recall

\begin{itemize}
\item {Critical energy density = $\rho_{crit}$ = $9\times10^{-10}$  J/m$^{3}$}. 
\item{Dark energy density = $\rho_{DE}$ = $0.70\times\rho_{crit}$ = $6.3\times10^{-10}$  J/m$^{3}$}
\item{$\hbar$ = $1.054\times 10^{-34}$ Js}
\item{G = $6.67\times10^{-11}$ m$^{3}$ kg$^{-1}$ s$^{-2}$}
\end{itemize}

\section{Conventional beliefs about the nature and investigation of dark energy }

Present conventional beliefs about dark energy density are that it is uniformly distributed in space and that its magnitude is given by $6.3\times10^{-10}$  J/m$^{3}$. The usual assumption is that every cubic meter of space contains the same dark energy density so that as the visible universe expands there is more total dark energy. I find it disqueting that most  physicists and astronomers are content to live with this violation of the conservation of energy, it leads to my doubts that the community has basic understanding of dark energy and has encouraged me to go in this new research direction.

$\rho_{DE}$ = $6.3\times10^{-10}$  J/m$^{3}$ initially strikes one as a very small energy density but as shown in the next section we experiment with smaller electric field energy densities in the laboratory.

\section{Comparison of dark energy density with the energy density of a weak electric field}

Consider a weak electric field $E$ = 1 volt/m. Using
\begin{equation}
\rho_{electric field} = \epsilon_{0}E^{2}/2 
\end{equation}

\begin{equation}
\rho_{electric field}=4.4\times10^{-12} \mbox{  J/m$^3$}
\end{equation}

\noindent Hence the energy density of this electric field is 100 times smaller than the dark energy density, $\rho_{DE}$ = $6.3\times10^{-10}$  J/m$^{3}$, yet this weak electric field is easily detected and measured. Thus we work with fields whose energy densities are much less than $\rho_{DE}$. This realization first started me thinking about the possibility of direct detection of dark energy

Of course, it is easy to sense and measure tiny electromagnetic fields; on the other hand there are obviously severe experimental problems in detecting dark energy density.
\begin{itemize}
\item{Unlike an electric field in the laboratory, we cannot turn dark energy on and off.}
\item{We do not know if there is a zero dark energy field to use as an experimental reference. In the fixed value, cosmological constant, explanation of dark energy, $\rho_{DE}$ has the same value in all space.}
\item{Even if the dark energy density should have a gradient, what force does it exert on a material object? }
\end{itemize}

\section{The terrestrial gravitational force field and a possible dark energy force}

In atom interferometry the phase change of atoms depends upon the integral of the potential difference between two separate trajectories of the atom in space. Of course at present we know nothing about whether or not dark energy exerts such a force. Indeed investigating this question is one of the purposes of our proposed experiment. In analogy we designate this force as $g_{DE}$ in units of force per unit mass.
 
Comments on $g$ and $g_{DE}$. 

\begin{enumerate}
\item{The gravitational force per unit mass on earth is $g$=9.8 m/s$^2$.}
\item{Atom interferometry studies have reached a sensitivity of much better than $10^{-9}$ $g$ in measurements of the gravitational acceleration \cite{chung} and found no anomaly. Even though a definite analysis for this has not be performed, it is probably safe to say that there is no evidence for $g_{DE}$ at this level.} 
\item{ Therefore $g_{DE}\leq10^{-8}$ m/s$^2$} using our assumptions about the properties of dark energy.
\end{enumerate}

\section{Preliminary considerations on how well we can null out $g$.}

Based on preliminary considerations we believe we can null out $g$ to a precision perhaps as small as $10^{-17}$. This sets the smallest $g_{DE}$ that we can investigate at $10^{-16}$ $m/s^2$.

\section{Assumptions about the properties of dark energy that make the experiment feasible}

We assume:

\begin{itemize}
\item {A dark energy force, $F_{DE}$, exists other than the gravitational force equivalent of $\rho_{DE}$.}
\item {$F_{DE}$ is sufficiently local and thus  $\rho_{DE}$ is sufficiently non-uniform so that $F_{DE}$ varies over a length of the order of a meter.}
\item {$F_{DE}$ acts on atoms leading to a potential energy $V_{DE}$.}
\item {The ratio $g_{DE}/g$ is large enough for $g_{DE}$ to be detected in this experiment by nulling signals from $g$.}
\end{itemize}

\section{Brief description of our experimental method}

The search for {$F_{DE}$ requires the nulling of all the known forces that can change the atomic phase. The effects of electric and magnetic forces are nulled by shielding and by using atoms in quantum states which are not sensitive to the linear Zeeman and Stark effects. The gravitational force is nulled by using two identical atom interferometers as described next.

\begin{figure}
\begin{center}
\includegraphics[width=4in]{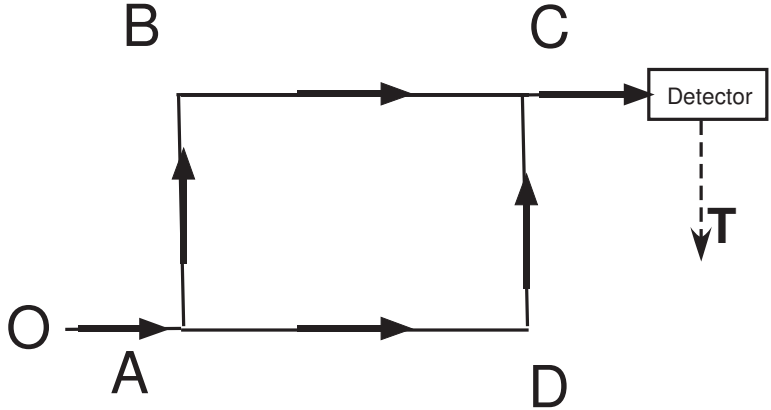}
\end{center}  
\caption{Diagram of an atom interferometer using the Mach-Zehnder concept.}
\end{figure}

In Fig. 1, a schematic diagram of an atom interferometer using the Mach-Zehnder concept, the solid lines represent atom beams. An atom beam from source O is split at A so that each atom quantum mechanically follows the two paths ABC and ADC. At D the two states arrive with relative phases, $\phi_{ABC}$ and $\phi_{ADC}$. The interference produces a signal $T$ proportional to the phase difference $\phi_{ABC}$-$\phi_{ADC}$. $T$ depends upon the potentials acting on the atoms in the space ABCD. The plane of the interferometer may be vertical or horizontal with respect to the earth's surface, our present preference is the vertical orientation. 

\begin{figure}
\begin{center}
\includegraphics[width=4in]{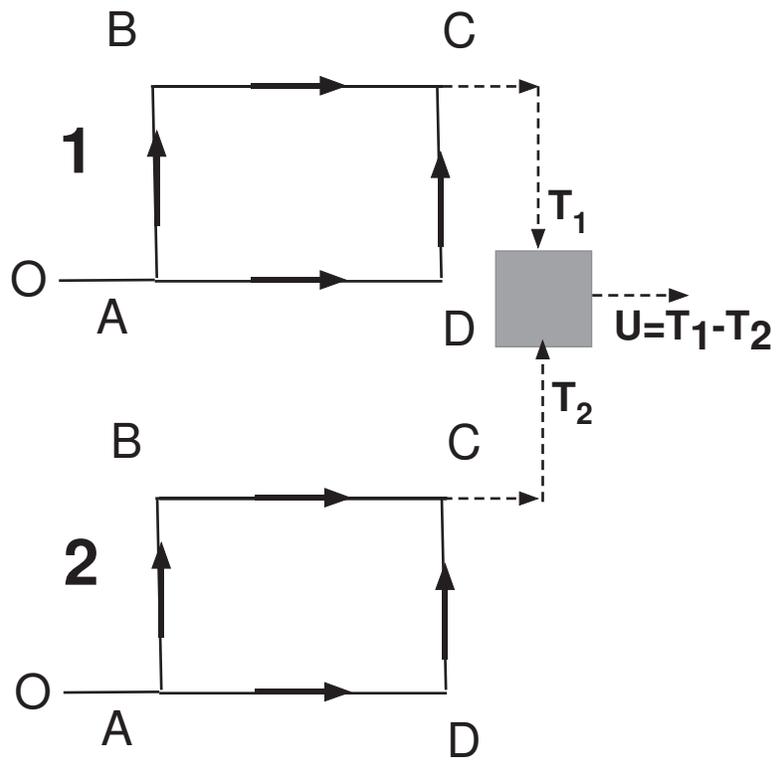}
\end{center}  
\caption{Diagram of the double atom interferometer.}
\end{figure}

Figure 2 shows the double apparatus schematically, two identical atom interferometers are used with the solid lines representing atom beams and the dash lines representing signal flow. The apparatus is in the vertical orientation. Interferometers 1 and 2 produce signals
 $T_{1}$ and  $T_{2}$, each signal being dependent on the potentials acting on the atoms in the spaces ABCD.  Considering just the earth's gravitational force $\vec{g}$, $T_{1}$ is proportional to the change in gravitational potential between the upper path ABC and the lower path ADC , and thus proportional to the gravitational acceleration $g$.

$U$ = $T_{1}$-$T_{2}$ is given by the difference between the accelerations of free fall at the locations of the first and second interferometers. If we assume $\vec{g}$ to be nearly constant at the earth's surface, $U$ = 0 for contributions from $\vec{g}$, except for small corrections. Thus signals from the gravitational force are nulled by this interferometer design. 

One realization of this design is a pair of fountain interferometers as described by Chung and his coworkers \cite{chung}. Even in a single interferometer, suppression of the signal due to $g$ to the $10^{-10}\,g$ level has already been demonstrated by subtracting a Newtonian model of tidal variations caused by the Moon, the Sun, and the planets. Using the pair of interferometers described above, we expect to be able to cancel the effects of gravity by a factor of $10^{-17}$.

\section{Nature of Sought Signal}

Figure 3 is a schematic illustration of how inhomogenious dark energy density could produce a non-zero signal $U$ with a value dependent upon the degree of inhomogeneity, the force exerted by dark enrgy on atoms, and the configuration of the double interferometer.

\begin{figure}
\begin{center}
\includegraphics[width=4in]{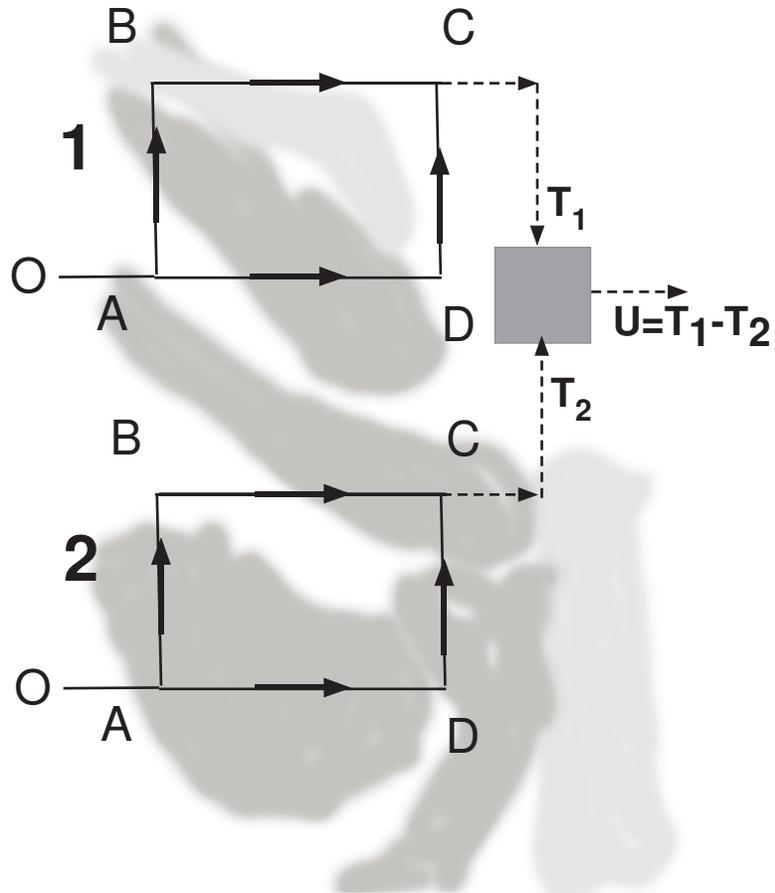}
\end{center}  
\caption{Illustration of how inhomogenious dark energy density could produce a non-zero signal. The gray shapes represent the assumed inhomogeneity of the distribution of dark energy density.}
\end{figure}

In this experiment the interferometers are fixed to the earth. The earth is spinning and moving in the Galaxy and the Galaxy is moving in the CMB frame with a velocity about 400km/s. Using present atom interfereometer readout methods , $U$ will be sampled at time intervals of the order of seconds to minutes. In this search dark energy density is assumed to be inhomogenious, but of course we know nothing about what the velocity of the dark energy density might be. In any case, the dark energy clumps are not tied to the earth. Therefore the sought signal will average over many samplings of different dark energy densities and will \emph{appear to be a noise signal}. This noise signal appearance has three consequences:

\begin{itemize}
\item {If a noiselike signal is found from output of the double interferometer, we must show tht it is not instrumental noise.}
\item {If a noiselike signal is found, we do not know how to show that it is related to dark energy}
\item {As noted in the next section the absence of a non-instrumental noiselike signal puts an upper bound on some other kinds of hypothetical forces and fields that might pervade the universe.}
\end{itemize}

\section{Other very weak forces and fields} 
My colleagues Holger Mueller and Ronald Adler have emphasized that this atom interferometry search is a general exploration of the possible existence of very weak forces, forces much weaker than gravity. Of course the criteria of inhomogeneity and an effect on matter must be met. 

Incidentally, to the best of my understanding, this research method is irrelevant to the grand old problem of understanding the cosmological significance of total zero-point vacuum energy.

\section{This experimental search method using an earth orbit satellite.}
Stern et al.\cite{stern} and Ertmer and Rasael \cite{ertmer} have emphasized the substantially increased precision obtained by carrying out atom interferometry experiments in the microgravity environment of an earth orbit satelite. There is a second advantage using an earth orbit satellite for atom interferometry searches for dark energy and other very weak forces. The nulling of $g$ is much easier.

\section{Acknowledgements}
I am deeply indebted to Holger Mueller and Ronald Adler for their patience in educating me in subjects ranging from the design of a a magentic optical trap to the proper application of general relativity to dark energy. I am grateful to my friends in the SLAC National Accelerator Laboratory for lunch table discussions and insightful criticisms of this speculative research.

I thank my partner in life, Joyce Beattie, for her wilingness to accept my preoccupation with this research.

This research is presently supported by Stanford University funds and by the author. The SLAC National Accelerator Laboratory is providing laboratory space and its continuing stimulating atmosphere.

\end{document}